\documentclass[english]{bvm2019}

\title{Open-source Tracked Ultrasound \\ with Anser EMT}

\author{Alfred~Michael~Franz$^{1,4}$, Herman~Alexander~Jaeger$^2$, Alexander~Seitel$^4$, P\'{a}draig~Cantillon-Murphy$^{2,3}$, Lena~Maier-Hein$^4$}

\authorrunning{Franz et al.}

\institute{
$^1$Institute for Computer Science, Ulm University of Applied Sciences, Ulm, Germany\\
$^2$University College Cork, Cork, Ireland\\
$^3$Tyndall National Institute, Dyke Parade, Cork, Ireland\\
$^4$Division of Computer Assisted Medical Interventions, German Cancer Research Center (DKFZ), Heidelberg, Germany\\
}

\email{franz@hs-ulm.de}

\begin{document}

%
\selectlanguage{english}

\maketitle

\begin{abstract}
Image-guided Interventions (IGT) have shown a huge potential to improve medical procedures or even allow for new treatment options. Most ultrasound(US)-based IGT systems use electromagnetic (EM) tracking for localizing US probes and instruments. However, EM tracking is not always reliable in clinical settings because the EM field can be disturbed by medical equipment. So far, most researchers used and studied commercial EM trackers with their IGT systems which in turn limited the possibilities to customize the trackers in order minimize distortions and make the systems robust for clinical use. In light of current good scientific practice initiatives that increasingly request research to publish the source code corresponding to a paper, the aim of this work was to test the feasibility of using the open-source EM tracker \textit{Anser EMT} for localizing US probes in a clinical US suite for the first time. The standardized protocol of Hummel~et~al. yielded a jitter of $0.1\pm0.1$~mm and a position error of $1.1\pm0.7$~mm, which is comparable to 0.1~mm and 1.0~mm of a commercial NDI Aurora system. The rotation error of Anser EMT was $0.15\pm0.16^{\circ}$, which is lower than at least $0.4^{\circ}$ for the commercial tracker. We consider tracked US as feasible with Anser EMT if an accuracy of 1-2~mm is sufficient for a specific application. 
\end{abstract} 

\section{Introduction}

Promising contributions in the area of Image-guided Interventions (IGT) \cite{2638-Cleary2010} have shown a huge potential to improve medical outcome of existing procedures or allow for new treatment options by enhancing the information available during the intervention. If, for example, the pose of an ultrasound (US) probe can be determined accurately, conventional US images can be enhanced in different ways: (1)~preoperative data can be visualized together with US images \cite{2638-Tomonari2013,2638-Franz2014}, (2)~instruments can be shown in relation to the US image \cite{2638-Tomonari2013}, and (3)~2D~US machines can record 3D images by combing 2D scans from different viewing angles \cite{2638-Mercier2005}. If localization data is used to train neural networks, 3D US is later possible without a tracker \cite{2638-Prevorst2017}.

Key component of many IGT systems is a tracking device, most frequently used for determining the pose of medical devices. Optical tracking allows for accurate localization \cite{2638-Cleary2010}, but requires a free line-of-sight (LoS) from a camera to optical markers which is cumbersome during freehand motion of a US probe. Electromagnetic (EM) trackers can localize small EM sensors in relation to a field generator (FG) without LoS \cite{2638-Franz2014}. Hence, most US-based IGT systems use EM tracking for localizing US probes (e.g., \cite{2638-Tomonari2013}). However, meanwhile it is apparent that EM tracking is not always reliable in clinical settings because the EM field can be disturbed, e.g. by medical devices or the patient stretcher \cite{2638-Franz2014}. To study these distortions, standardized assessment protocols for testing EM trackers in specific clinical environments with a maximum of comparability have been published \cite{2638-Franz2014,2638-Hummel2005}. So far, most researchers used and studied commercial EM trackers with their IGT systems which in turn limited the possibilities to customize the trackers in order minimize distortions and make the systems robust for clinical use.

In parallel, recent discussions in the scientific community yielded the request to publish all source code of scientific results \cite{2638-Ince2012}. In the special case of IGT prototypes, this should at best include the source code for the tracking algorithms. In this regard, a welcome development is that open-source EM tracking systems have been published recently \cite{2638-Li2014,2638-Jaeger2017} and enable researchers to develop IGT systems with open soft- and hardware. However, these systems have not been tested in a tracked US context so far and it remains unclear if tracking is accurate and robust in related clinical environments. In this study, we assess the feasibility of using the open-source EM tracker \textit{Anser EMT} \cite{2638-Jaeger2017} for localizing US probes in a clinical US suite for the first time.

\section{Materials and Methods}

\subsection{Tracked Ultrasound Setup}
\begin{figure}[t]
	\centering
	\includegraphics[width=0.572\textwidth]{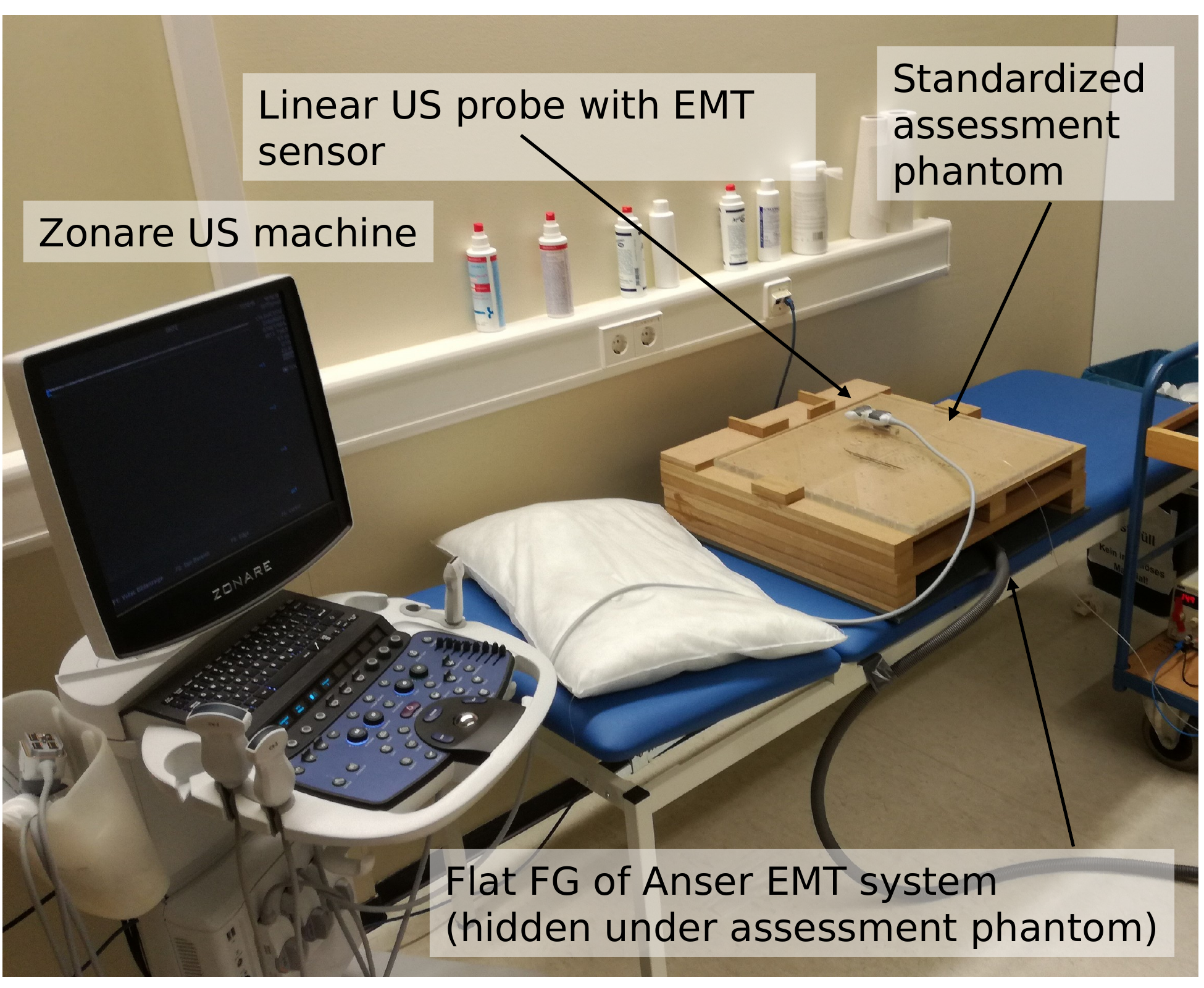} 
	\includegraphics[width=0.35\textwidth]{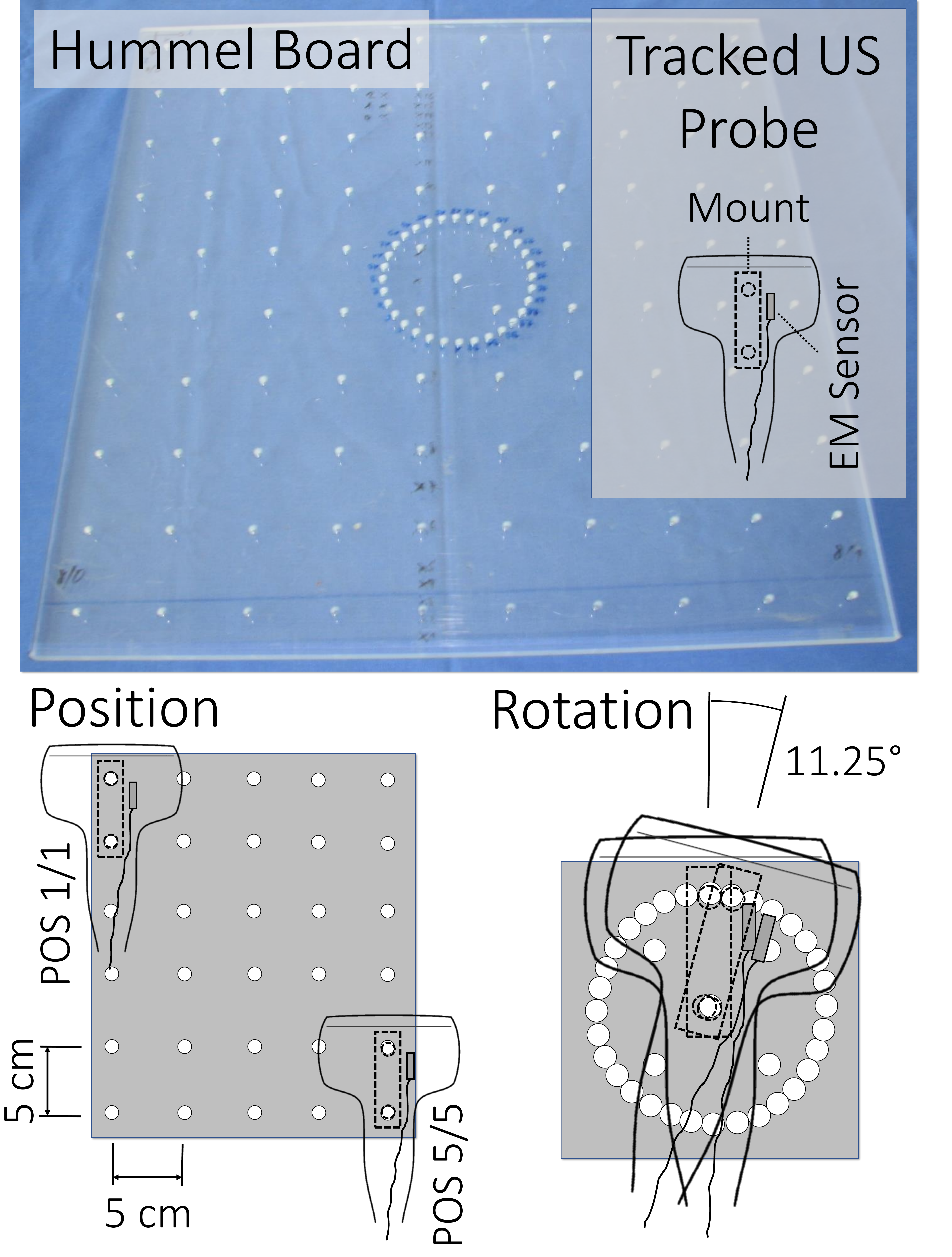} 
	\caption[]{Experimental setup in the US suite. A linear US probe is equipped with an EMT sensor and fixed on a special mount. The mount can be moved to known positions on the standardized assessment phantom (Hummel Board).}
	\label{fig:Setup}
\end{figure}

Anser EMT is a open-source EM tracking platform for IGT \cite{2638-Jaeger2017}. Full system design schematics, instructions, and code can be accessed online (\url{http://openemt.org}). The flat FG of Anser EMT creates a magnetic field in a working volume of 25~x~25~x~25~$cm^3$ and is capable of tracking up to 16 EM sensors in the latest version. In this study, a NDI~5-DOF sensor (Northern Digital Inc.(NDI), Waterloo, Canada, Model no. 610099) was used. It was fixed to a linear US probe (type L14-5w) of a Zonare ZS3 Ultrasound System (ZONARE Medical Systems (inc), Mountain View, California) as shown in Fig.~\ref{fig:Setup}. 

Anser EMT supports the OpenIGTLink \cite{2638-Tokuda2009} protocol for connection to open-source IGT toolkits. For this study, the Medical Imaging Interaction Toolkit (MITK) was connected to the system and the plugin \textit{Hummel Protocol Measurements}\footnote{org.mitk.gui.qt.igt.app.hummelprotocolmeasurements} was used for further processing and evaluation of the data. All software used for this project is available open-source under the link mentioned earlier and in the MITK repository\footnote{\url{https://phabricator.mitk.org/source/mitk/}}.

\subsection{Standardized Assessment of Tracked Ultrasound}
We used a standardized assessment protocol proposed by Hummel~et~al. \cite{2638-Hummel2005} to assess our tracked US setup. 5~x~5~=~25 positional measurements were performed on a polycarbonate board (Hummel Board) in a known grid with 5~cm distances as shown in Fig.~\ref{fig:Setup}. Orientational measurements were done in 31 steps of 11.25$^\circ$ for a 360$^\circ$ rotation the middle of the board. For all positions, 150 measurement samples were recorded over 10~s at an update rate of 15 Hz. 

The jitter error at one position was defined as the root mean square error of 150 samples. To determine positional accuracy, the Euclidean distances between two adjacent measured sensor locations, each averaged over 150 samples, were computed. The deviation to the reference of 5~cm was defined as distance error and determined for all 16 distance measurements (4 horizontal x 4 vertical) of the 5~x~5 grid. As another measure for positional accuracy, a grid matching error was determined. This error represents the fiducial registration error (FRE) obtained when matching the measured grid positions (n=25) to the grid of known reference positions with the optimal rigid transformation in a least square sense \cite{2638-Horn1987}. The angle differences between pairs of measured orientations and the known relative sensor rotation of 11.25$^{\circ}$ were determined to get the orientational errors.

The assessment of the tracked US with Anser EMT was performed in a clinical US suite of the German Cancer Research Center (DKFZ), as shown in Fig.~\ref{fig:Setup}. The Hummel Board was placed at a height of 11~cm above the FG on the patient stretcher. The FG was aligned in the middle of the covered volume. For comparison, the position measurements were repeated in the same setup on the patient stretcher but without a US probe (US suite). In addition, reference data of comparable experiments (position and rotation) in a distortion-free lab environment (Lab ref) is available from a previous study \cite{2638-Jaeger2017}.

\section{Results}

\begin{SCfigure}[5][t]
	\label{fig:boxplots}
	\setlength{\figbreite}{0.2\linewidth}
	\caption{Relative position errors of 4~x~4~=~16 measured 5~cm~distances illustrated as box-whiskers plots. The diamonds show the mean values, the whiskers the minimum and maximum values.}
	\includegraphics[height=0.5\textwidth,angle=-90]{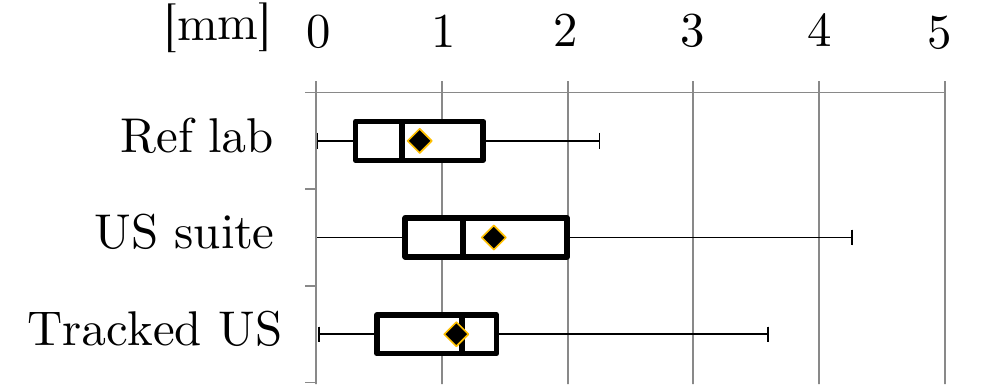} 
\end{SCfigure}

\begin{table}[b]
\caption{Comparison to the NDI Aurora tracker in the US suite \cite{2638-Franz2013}. A subset of 4~x~3~=~12 positions and 180$^{\circ}$ of rotation measurements of the Anser EMT data was evaluated for this table to be comparable. Note, that the configuration of the field generator (FG) in \cite{2638-Franz2013} was different. We took the results from the bottom level of \cite{2638-Franz2013} which had a similar distance to patient stretcher and FG as in this study.}
\label{2638-tab-comp}
\begin{tabular*}{\textwidth}{l@{\extracolsep\fill}lp{3.2cm}lll}
\hline
System & Setup & Field Generator & Prec.[mm] & Acc.[mm]& Rot\_1[$^{\circ}$] \\
\hline
Anser EMT & Tracked US & Flat FG (below) & 0.07 & 1.11 & 0.1 \\
Anser EMT & US Suite & Flat FG (below) & 0.18 & 1.65 & <na>\\
NDI Aurora& US Suite & Compact FG with US probe (above) \cite{2638-Franz2012}  & 0.09 & 1.03 & 0.4 \\	
\hline
\end{tabular*}
\end{table}

The precision (jitter error) was $0.1\pm0.1$~mm (Lab and Tracked US) and $0.2\pm0.1$~mm (US Suite) on average~($\mu\pm\sigma$, n=25 grid points) with a maximum error of 0.7~mm. The positional errors of the 5~cm distance measurements on the board in all setups are shown as boxplots in Fig.~\ref{fig:boxplots} and usually stay below 2~mm. The grid matching error was 1.5~mm (Lab), 2.2~mm (US suite) and 2.9~mm (Tracked US).
The average sensor locations in the tracked US setup are visualized in Fig.~\ref{fig:Results}. Orientation measurements yielded an error of $0.15\pm0.16^{\circ}$ (n=31) in the tracked US setup which was increased by around $0.1^{\circ}$ compared to reference measurement ($0.04\pm0.02^{\circ}$ \cite{2638-Jaeger2017}). All measurements taken in this study are provided open data in the Open Science Framework (\url{https://osf.io/aphzv/}) together with comparative data sets of other trackers.

\section{Discussion}

\begin{figure}[t]
	\centering
	\includegraphics[width=1.0\textwidth]{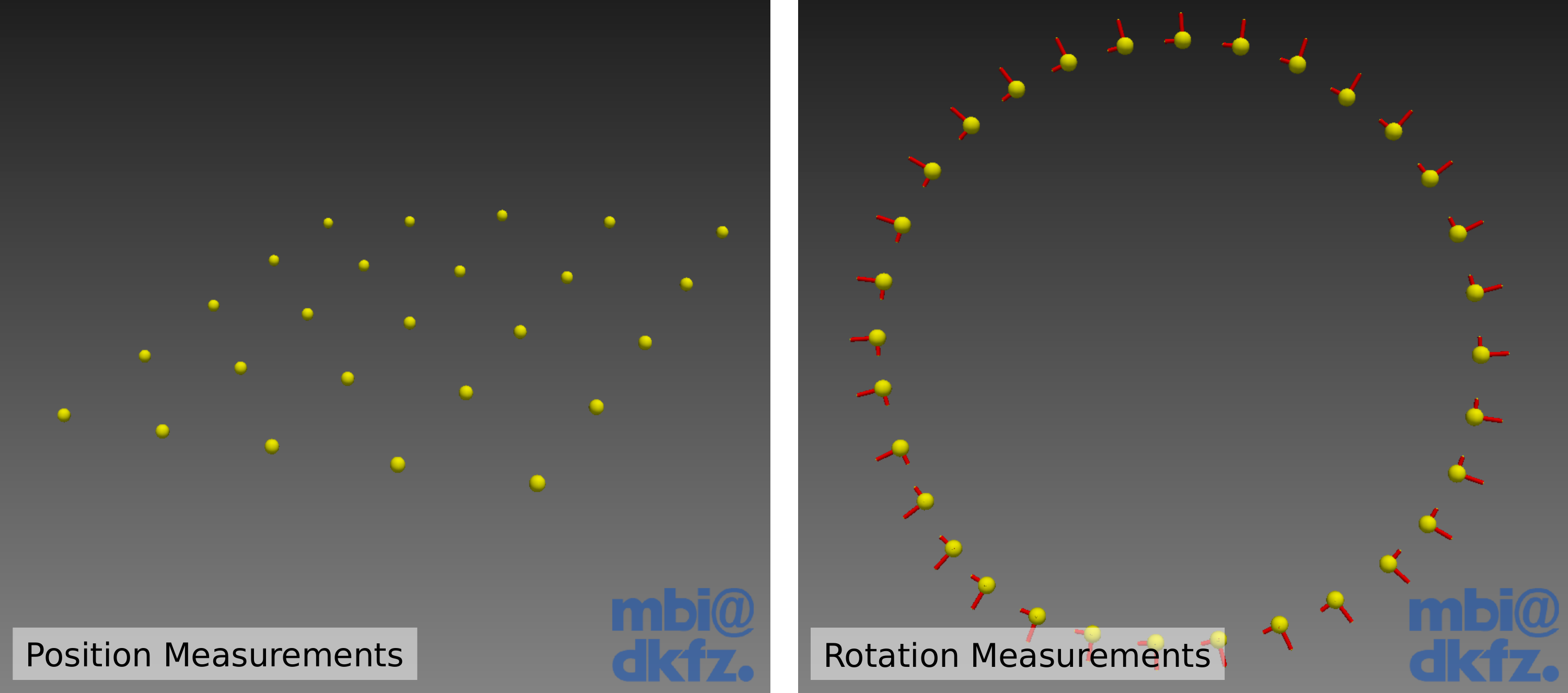} 
	\caption[]{3D visualization of the measurements in the Tracked US setup. Left: 25 grid points of the position measurements. Only a slight distortion in the middle of the back row is visible. Right: 32 rotation measurements, visualized as the measured position together with the sensor coordinate axes in red. The circle shows no visual distortion of the measurements. Please note that only the sensor axis (longer red line) was clearly defined because a 5 DoF sensor was used.}
	\label{fig:Results}
\end{figure}

The jitter error of $0.1\pm0.1$~mm and the position error of $1.1\pm0.7$~mm is comparable to 0.1~mm jitter and 1.0~mm position error of a commercial tracker (NDI Aurora) in the same environment \cite{2638-Franz2013} as shown in Table~\ref{2638-tab-comp}. As for the measurements in a laboratory environment in an earlier study \cite{2638-Jaeger2017}, the rotational errors of Anser EMT are also small, below $0.3^{\circ}$ in our measurements, which is better than published results of other trackers (e.g., at least $0.4^{\circ}$ \cite{2638-Franz2013}, but up to $3^{\circ}$ \cite{2638-Franz2012} for a NDI Aurora system). When looking at the grid matching error, we see a slight drop in accuracy between US~suite~(2.2~mm) and Tracked~US~(2.9~mm). This is not reflected by the 5~cm distance evaluation, where the median error is similar in both setups (1.2~mm). If only distances between pairs of points are evaluated, slight field distortions as we see in the back row of the position measurements (cf. Fig.~\ref{fig:Results}) can have little effect on the metric, but matching the whole grid can reveal these distortions. Therefore, we propose to always have a look at both metrics, and also at the raw data points, when interpreting Hummel protocol results.

Most errors are relatively small, which is good for the system, but raises the question if manual measurements are accurate enough to determine its limits in accuracy. In an earlier study a reference measurement was repeated three times to analyze reproducibility \cite{2638-Franz2012}. The average 5~cm distance error measured was in the range of 0.3-0.5~mm. In case of this study, the difference between the average errors in the US suite (1.4~mm) and Tracked US (1.1~mm) setup, as shown in Fig.~\ref{fig:boxplots}, might be caused by the natural variation of manual measurements. However, the results still show, that the errors stay below 1 to 2 mm in most of the cases and demonstrate a high accuracy for the tracked US setup.

We used a 5 DoF sensor for our experiments. Depending on the application, 6 DoF of the probe are required. In this case either a second 5 DoF sensor or a slightly bigger 6 DoF sensor can be used. According to our experience with EM trackers, it is unlikely that a second or different sensor would affect tracking accuracy or robustness except for slight manufacturing tolerances.

All in all, although the positional error is slightly increased when tracking the US probe in the US suite, we consider tracked US as feasible with the Anser EMT system if an accuracy of 1-2~mm is sufficient for a specific application. 

\section*{Acknowledgements}
This work was supported by the European Union through the ERC starting Grant COMBIOSCOPY (ERC-2015-StG-37960) and by the Irish Health Research Board (POR/2012/31), Science Foundation Ireland (15/TIDA/2846). We acknowledge the German Cancer Research Center (DKFZ), Heidelberg, and the Institute of Image-Guided Surgery (IHU), Strasbourg for supporting this work.

\bibliographystyle{bvm2019}

\bibliography{2638}
\end{document}